\documentstyle[12pt,psfig]{article}
\def\duetre{{\textstyle {{\rm 2}\over {\rm 3}}}}

\def\eq{\begin{equation}}
\def\ee{\end{equation}}
\def\eqa{\begin{eqnarray}}
\def\eea{\end{eqnarray}}

\def\bra#1{\mbox{$\langle #1\vert $}}
\def\ket#1{\mbox{$\vert #1\rangle$}}

\def\bold#1{\setbox0=\hbox{$#1$}%
     \kern-.025em\copy0\kern-\wd0
     \kern.05em\copy0\kern-\wd0
     \kern-.025em\raise.0433em\box0 }                 

\def\bi#1{\setbox0=\hbox{$#1$}%
     \kern-.025em\copy0\kern-\wd0
     \kern.05em\copy0\kern-\wd0
     \kern-.025em\raise.0433em\box0 }                 

\begin{document}


\centerline{\Large{\bf Nucleon form factors in a chiral constituent-quark
model}}

\vskip 0.7cm

\centerline{S.~Boffi$^{a}$, P.~Demetriou$^{a}$, M.~Radici$^{a}$ and
R.~F.~Wagenbrunn$^{b}$}

\medskip

\centerline{\footnotesize \sl $^a$~Dipartimento di Fisica Nucleare e Teorica,
Universit\`a di Pavia and INFN, Pavia, Italy}

\centerline{\footnotesize \sl $^b$~Institute for Theoretical Physics,
University of Graz, Graz, Austria}

\vskip 1.5truecm


\begin{abstract}

\noindent The electromagnetic form factors of the nucleon have been calculated
in a chiral constituent-quark model. The nucleon wave functions are obtained by
solving a Schr\"odinger-type equation  for a semi-relativistic Hamiltonian
with an effective interaction derived from the exchange 
of mesons belonging to the pseudoscalar octet and singlet and a linear
confinement potential. The
charge-density current operator has been constructed consistently 
with the model Hamiltonian in order to
preserve gauge invariance and to satisfy the continuity equation.

\end{abstract}

\bigskip




\begin{section}{Introduction}

For a long time constituent-quark models (CQM) have been proposed to
explain spectroscopic properties of hadrons within a nonrelativistic
framework~\cite{[IK]} (see also
refs.~\cite{[Karl],[Giannini]} and references therein). In
these models the effective degrees of freedom are massive quarks moving in a
long-range confinement potential with
the basic SU(6) spin-flavour symmetry. According to the analysis of
ref.~\cite{[DeRujula]} the residual interaction responsible for the SU(6)
symmetry breaking is described by the one-gluon
exchange diagram and is identified with the hyperfine-like part of its
nonrelativistic reduction. Given the fact that the mass of the three
constituent quarks is small, the nonrelativistic approximation is not {\sl a
priori} justified. In addition some relativistic corrections are
also necessary to account for the observed small
spin-orbit effects~\cite{[Godfrey],[CI]}. Relativized versions of CQM were then
discussed~\cite{[Warns],[Capstick]}. Alternatively, relativity is considered
from the very beginning in the CQM adopting the light-front
formalism~\cite{[Cardarelli],[Cardarellib]} or using the socalled
Bakamjian-Thomas construction~\cite{[BT]} to derive the Poincar\'e invariant
formulation of the quark model for baryons~\cite{[Cotanch]}.

While rather successful in describing the octet and decuplet ground states,
these models 
still face some problems, such as, e.g., the wrong level orderings of
positive- and negative-parity excitations, which can be traced back to
inadequate symmetry properties of the one-gluon-exchange interaction.

The existence of an increasing number of near-parity doublets in the high-energy
sector suggests that the approximate chiral symmetry of quantum chromodynamics
(QCD) is realized in the hidden Nambu-Goldstone mode at low excitation and in
the explicit Wigner-Weyl mode at high excitation~\cite{[Manohar]}.
Thus the spontaneous breaking of chiral symmetry and the associated appearence
of the octet of pseudoscalar mesons as the approximate Goldstone bosons induces
a chiral interaction between quarks that is mediated by such
mesons~\cite{[Brauer]}. Its spin and flavour dependence
modifies the symmetry properties of the Hamiltonian and ultimately leads to a
correct ordering of the positive- and negative-parity states in the baryon
spectra~\cite{[Glozman]}.

Various hybrid models have been constructed advocating meson exchanges in
addition to sizeable contributions coming from gluon exchanges (see, e.g.
refs.~\cite{[Obukhovsky],[BHF],[valcarce],[fabre],[Shen]}).

Recently, a chiral CQM has been proposed whose effective quark-quark
interaction is derived from Goldstone-boson exchange alone involving the
pseudoscalar meson octet and singlet~\cite{[Plessas],[Plessasb]}. The model is
capable of providing a unified description not only of the nucleon and Delta
spectra but also of all strange baryons.

A stringent test of the model would be to probe its eigensolutions in the
description of the electromagnetic properties of baryons. In this paper the
nucleon electromagnetic form factors are calculated without free parameters
starting from the nucleon wave functions obtained with the model of
refs.~\cite{[Plessas],[Plessasb]}
and using a charge-current density operator consistently
derived along the lines proposed in ref.~\cite{[Capuzzi]}. The model is
briefly reviewed in section 2, while the expression of the charge-current
density operator is given in sect. 3. The results are presented and discussed
in sect. 4. 

\end{section}

\begin{section}{The model}

The chiral model of refs.~\cite{[Plessas],[Plessasb]} 
is semi-relativistic in the sense that the
kinetic energy operator is taken in relativistic form:
\eq 
H_0 = \sum_{i=1}^3 \sqrt{{\vec p}^2_i + m_i},
\label{eq:kinetic}
\ee
with $m_i$ the masses and ${\vec p}_i$ the three-momenta of the constituent
quarks. This form ensures the average quark velocity to be lower than the
light velocity, a requirement that is usually not fulfilled by nonrelativistic 
models.
In addition, by the choice (\ref{eq:kinetic}) one excludes negative-energy
states {\sl ab initio\/} and simply solves a Schr\"odinger equation for bound
states without facing all the complications of a fully covariant treatment of
the three-quark system.

The dynamical part consists of a linear confinement potential, 
\eq
V_{\rm conf} ({\vec r}_{ij}) = V_0 + C r_{ij}
\ee
depending on the interquark distance $r_{ij}$ and the two fitting parameters
$V_0$ and $C$, and a sum of pseudoscalar meson-exchange potentials:
\eqa
V_\chi^{\rm octet} ({\vec r}_{ij}) 
&=& \Big[
\sum_{a=1}^3 V_\pi({\vec r}_{ij})\lambda_i^a\lambda_j^a +
\sum_{a=4}^7 V_K({\vec r}_{ij})\lambda_i^a\lambda_j^a \nonumber \\
&{}&\qquad{} +
 V_\eta({\vec r}_{ij})\lambda_i^8\lambda_j^8 \Big]
{\vec\sigma}_i\cdot{\vec\sigma}_j 
\label{eq:octet} \\
&{}&\nonumber \\
V_\chi^{\rm singlet} ({\vec r}_{ij}) 
&=& \duetre\, V_{\eta'}({\vec r}_{ij})\,
{\vec\sigma}_i\cdot{\vec\sigma}_j, 
\label{eq:singlet} 
\eea
where ${\vec\sigma}_i$ and ${\vec\lambda}_i$ are the quark spin and flavour
matrices, respectively. In the static approximation used in
refs.~\cite{[Plessas],[Plessasb]}, the meson-exchange potentials are given by
\eq
V_\gamma ({\vec r}_{ij}) = {g^2_\gamma\over 4\pi}{1\over 12\, m_i\,m_j}
\left[\mu_\gamma^2\,{{\rm e}^{-\mu_\gamma r_{ij}}\over r_{ij}} 
- 4\pi\,\delta({\vec r}_{ij})\right],
\label{eq:static}
\ee
with $\mu_\gamma$ being the meson masses and $g_\gamma$ the meson-quark
coupling constants ($\gamma = \pi, K, \eta,\eta'$). In the chiral limit there is
only one coupling constant $g_8$ for all Goldstone bosons. Due to the special
character of the singlet $\eta'$ meson, its coupling constant $g_0$ was allowed
to deviate from $g_8$.

Since one deals with structured particles (constituent quarks and mesons) of
finite extension, one has to smear out the $\delta$ function in eq.
(\ref{eq:static}). In refs.~\cite{[Plessas],[Plessasb]} 
a Yukawa-type smearing was used,
i.e. \eq 4\pi\,\delta({\vec r}_{ij}) \longrightarrow \Lambda_\gamma^2 \,
{{\rm e}^{-\Lambda_\gamma r_{ij}}\over r_{ij}},
\ee
involving the cutoff parameters $\Lambda_\gamma$, which were assumed to follow
a linear scaling with meson masses:
\eq
\Lambda_\gamma = \Lambda_0 + \kappa\mu_\gamma.
\ee

Once the quark masses are fixed, the model has five fitting parameters, i.e. the
depth $V_0$ and the slope $C$ of the confinement potential, the ratio
$g_0/g_8$ of the singlet to octet meson-quark coupling constants and the two
parameters $\Lambda_0$ and  $\kappa$ defining $\Lambda_\gamma$.

The specific spin-flavour symmetry inherent in the chiral potential is
responsible for the correct level structure. According to the selected values
of the quark masses different sets of numerical values for the five free
parameters can produce fits to the baryon spectra of comparable
quality.

The Schr\"odinger-type equation for the model is 
accurately solved in the stochastic
variational method~\cite{[Varga]} using wave functions that are expanded in
basis functions involving correlated Gaussians as follows
\eqa
\Phi^k_{JM,TM_T}({\vec x}_k,{\vec y}_k) 
&=& x_k^\lambda\, y_k^l\, \exp(-\beta x^2_k - \delta y^2_k + \gamma {\vec
x}_k\cdot{\vec y}_k) \nonumber \\
& &\times \left[{\cal Y}^L_{\lambda l}({\hat x}_k,{\hat y}_k) \bigotimes
\chi^k_{(s_0,1/2)S}\right]_{JM}\,\chi^k_{(t_0,1/2)TM_T}, 
\label{eq:basis}\\  \nonumber
\eea
depending on the Jacobi coordinates of partition $k$,
\eqa
{\vec x}_k &=& {\vec r}_p - {\vec r}_q , \nonumber\\
{\vec y}_k &=& {\vec r}_k - \frac{m_p {\vec r}_p + m_q {\vec r}_q}{m_p + m_q},
\\  \nonumber
\eea
where ${\vec r}_i$ and $m_i$ ($i=1,2,3$) are the particle coordinates and
masses, and ($k,p,q$) is an even permutation of ($1,2,3$). The 
${\cal Y}^L_{\lambda l}$ represent the bipolar spherical harmonics
\eq
{\cal Y}^{LM_L}_{\lambda l} ({\hat x}_k,{\hat y}_k)
= \left[ Y_\lambda({\hat x}_k)\bigotimes Y_l({\hat y}_k)\right]_{LM_L} .
\ee
The spin (isospin) parts arise from coupling single-particle spins (isospins)
following the scheme
\eqa
\chi^k_{(s_0,1/2)SM_S} &=& \left[\chi^{(pq)}_{(1/2,1/2)s_0} \bigotimes
\chi_{1/2}^k\right]_{SM_S} , \nonumber\\
\chi^{(pq)}_{(1/2,1/2)s_0m_0} &=&
\left[\chi^p_{1/2}\bigotimes\chi^q_{1/2}\right]_{s_0m_0} .\\
\nonumber
\eea
For a given total angular momentum $J$ and isospin $T$ the stochastic
variational method selects basis functions according to a set of 6 discrete
parameters ($L,\lambda,l,S,s_0,t_0$) and 3 continuous parameters
($\beta,\gamma,\delta$). $L$ is the total orbital angular momentum, $\lambda$
and $l$ are the orbital angular momenta corresponding to ${\vec x}_k$ and
${\vec y}_k$, $S$ is the total spin, $s_0$ the spin and $t_0$ the isospin of
the subsystem ($pq$).

The total wave function is composed of a symmetrized linear combination of basis
wave functions of the form (\ref{eq:basis}). 

The Schr\"odinger equation can also be solved in momentum space. The basis
functions are of the same form as in eq. (\ref{eq:basis}) in terms of the
corresponding Jacobi conjugate momenta (${\vec p}_{x_k},{\vec p}_{y_k}$).
The method has the clear advantage of producing analytical solutions for the
total baryon wave function both in momentum and space coordinates.

\end{section}

\begin{section}{The charge-current density operator}

The relativistic form of the kinetic energy does not permit the use of the
traditional one-body current density operator, nor is it necessary to adopt
sophisticated procedures to include relativistic effects, such as those
proposed, e.g., in
refs.~\cite{[Warns],[Cardarelli],[CloseLi],[CK],[Santopinto]}.
A gauge invariant charge-current density operator can be derived consistently
with the model Hamiltonian. It contains both one- and two-body terms. The
one-body contribution includes the charge, the convective- and the spin-current
operators. Their matrix elements between free particle states are obtained
following the functional derivative formalism proposed in
ref.~\cite{[Capuzzi]}. For a particle of charge $e$ and mass $m$ they are given
in momentum space by the following expressions, respectively: \eqa
\bra{{\vec p}'}j_0({\vec x})\ket{{\vec p}} 
&=& e \, {1\over (2\pi)^{3}}\, {\rm e}^{{\rm i}({\vec p}-{\vec p}')\cdot{\vec x}},
\label{eq:charge}\\
& &\nonumber \\
\bra{{\vec p}'}{\vec j}({\vec x})\ket{{\vec p}} 
&=& e {{\vec p} + {\vec p}'\over  E_{\vec p} + E_{{\vec p}'}}
\,{1\over (2\pi)^{3}}\, {\rm e}^{{\rm i}({\vec p}-{\vec p}')\cdot{\vec x}}, 
\label{eq:convective}\\
& &\nonumber \\
\bra{{\vec p}',s'} {\vec j}^S({\vec x})\ket{{\vec p},s} 
&=& {{\rm i} e\over E_{\vec p} + E_{{\vec p}'}}
 \bra{s'}{\vec\sigma}\times({\vec p}'-{\vec p})\ket{s}
\,{1\over (2\pi)^{3}}\, e^{{\rm i}({\vec p}-{\vec p}')\cdot{\vec x}}, 
\label{eq:spin}\\ \nonumber
\eea
where
\eq
E_{\vec p} = \sqrt{{\vec p}^2 + m^2} , \qquad
 E_{{\vec p}'} = \sqrt{{\vec p}'{}^2 + m^2}.
\ee

With respect to the usual nonrelativistic expressions, in the 
semi-re\-la\-ti\-vis\-tic
approach only the spatial components of the charge-current density operator are
affected, while the time component is simply given by the charge density.
Furthermore, the energy denominator appearing in the current matrix elements
reduces to $2m$ in the low-energy limit recovering the nonrelativistic
approximation. 

The two-body current operator can be derived directly from the continuity
equation consistently with the model Hamiltonian described in sect. 2 (see also
ref.~\cite{[Glozman]}). It is given by
\eqa
{\vec j}_{\rm ex}({\vec p}_1,{\vec p}_2) &=& -{\rm i} \,e\,
 \sum_\gamma
\left\{ \left[ {\vec\tau}^{(1)}\times{\vec\tau}^{(2)}\right]_3
\delta_{\gamma\pi} +  \left(\lambda^{(1)}_4  \lambda^{(2)}_5 -\lambda^{(1)}_5
\lambda^{(2)}_4 \right)\,\delta_{\gamma K} \right\} \nonumber \\
& &\nonumber \\
& &\ \, \times \left\{ 
 {g_\gamma^2\mu_\gamma^2\over 12\,m_1\,m_2} 
{{\vec p}_2 -{\vec p}_1\over (\mu_\gamma^2 + p_1^2)(\mu_\gamma^2 + p_2^2) }
- \left(\mu_\gamma \rightarrow \Lambda_\gamma\right) \right\}
{\vec\sigma}^{(1)}\cdot{\vec\sigma}^{(2)}. \nonumber\\
& &
\eea
The $\lambda_{4,5}$ matrices are related to the SU(2) V-spin subgroup
contained within SU(3). Together with the exchange character of the isospin
dependence of the two-body current they result from the commutator between 
the quark-quark potential and the charge density that depends on $\lambda_3$ and
$\lambda_8$. Therefore only pion and kaon exchanges are allowed.

\end{section}

\begin{section}{The nucleon form factors}

As a first test of the model the electromagnetic form factors of the nucleon
have been calculated. In this case only the one-pion exchange part of the
two-body interaction, eq.~(\ref{eq:octet}), contributes. The calculation is
fully consistent and without free parameters. In the following we give results
for two parametrizations of the chiral constituent-quark model with
different values for the constituent-quark masses. The first parametrization
is from refs.~\cite{[Plessas],[Plessasb]} with constituent-quark masses
$m_{u,d} = 340$ MeV. The second is a modified version with $m_{u,d} =250$
MeV and accordingly readjusted parameters of the model in order to obtain
baryon spectra of similar quality. 

The electric ($G_E$) and magnetic ($G_M$) form factors are plotted in fig. 1 for
the proton and in fig. 2 for the neutron. The thin (thick) solid lines refer to
$m_{u,d}=$ 250 (340) MeV. 

Two remarks can be made on the results. First, the fall-off of $G_E^p$ and
$G_M^{p,n}$ as a function of $Q^2$ is lower than observed. This $Q^2$
dependence reflects the fact that here the constituent quarks are assumed to be
point-like. As in other constituent-quark models this assumption underestimates
the electromagnetic radii of the nucleon. Second, the value of $G_M^{p,n}$ at
$Q=0$ does not reproduce the nucleon magnetic moment. However, the ratio
$G_M^p/G_M^n$ is in good agreement with the corresponding observed ratio of the
proton to neutron magnetic moment, a feature common to all nonrelativistic
constituent-quark models. The discrepancy at $Q=0$ is due to two effects.
(a) Two-body currents do not contribute to the nucleon magnetic form factor
because the Hamiltonian of eqs.~(\ref{eq:octet}) and (\ref{eq:singlet}) does not
contain the full axial dipole-dipole interaction that describes one-pion
exchange completely. Therefore it is not possible to mix different values of the
orbital angular momentum with the same parity. (b) The semi-relativistic
form of the one-body current with an energy-dependent denominator suppresses its
contribution with respect to the nonrelativistic case; it would therefore
require a rather lower value for the constituent-quark mass in order to 
reproduce the nucleon magnetic moments.  

In order to improve the quality of the results without destroying the
agreement with the observed baryon spectra one has to consider that
constituent quarks are effective degrees of freedom with some spatial 
extension~\cite{[Vogl],[Cardarelli]}.
As such, a charge form factor $f(Q^2)$ could be appended to the charge density
operator (\ref{eq:charge}) and the convective part of the current density
operator (\ref{eq:convective}) as well as a magnetic form factor $g(Q^2)$ to the
spin part of the current density operator (\ref{eq:spin}). This will modify the
$Q^2$ dependence. In fact, a rather good agreement with data can already be
obtained for $G_M^{p,n}$ at $Q^2> 0.5$ (GeV/$c$)$^2$ assuming a simple dipole
form factor 
\eq
f(Q^2) = {1\over [1 + a Q^2]^2}
\label{eq:dipole}
\ee 
common to all($u$ and $d$) quarks. This is achieved with a rather small quark charge radius,
i.e. $r_c = 0.35$ fm, almost independently of its mass. Due to the small
radius $r_c$ the $Q^2$ dependence of $G_E^p$, although largely improved, could
not yet be reproduced.

On the other hand, once constituent quarks are treated as extended
objects, it is not unreasonable to introduce an anomalous magnetic moment
$\kappa$ in their electromagnetic form factor. Thus, besides a dipole form for
$f(Q^2)$, the following form for $g(Q^2)$ has been considered:
\eq
g(Q^2) = f(Q^2) + \kappa {1\over [1 + b Q^2]^3} .
\label{eq:cappa}
\ee
The actual value of $\kappa$ has been fixed in order to obtain the
experimental value of the proton magnetic moment. For a quark mass $m=340$
(250) MeV one obtains $\kappa=0.867$ (0.549). Correspondingly, the neutron
magnetic moment turns out to be $-1.828$ ($-1.812$) n.m. in good agreement with
experiment. The other two parameters $a$ and $b$ in eqs. (\ref{eq:dipole}) and
(\ref{eq:cappa}) are then fixed by fitting the $Q^2$ dependence of $G_M^p$. The
resulting values for the quark charge and magnetic radius are: $r_c = 0.691$ fm
and $r_m = 1.050$ (0.935) fm with a quark mass $m=340$ (250) MeV. It is worth
noting that the extracted value of the quark charge radius is significantly
close to the value required by the assumption of vector-meson dominance. Without
any free parameter one can then calculate the other nucleon form factors. The
results are shown in figs. 1 and 2 by the thin (thick) dashed lines for $m=250$
(340) MeV. A rather satisfactory agreement is obtained, especially for $G_E^p$.

The electric form factor of the neutron turns out to be too small in all cases.
This is due to the deficiency of the charge density operator
(\ref{eq:charge}) derived in ref.~\cite{[Capuzzi]}. In order to be consistent
with the semi-relativistic Hamiltonian no other contributions involving 
spin-dependent terms, like e.g. the Darwin-Foldy correction, are possible and 
the quark charges add up to a total vanishing neutron charge. 

\end{section}

\begin{section}{Conclusions}

A completely consistent calculation of the nucleon electromagnetic form
factors has been performed within the chiral constituent-quark model proposed in
refs.~\cite{[Plessas],[Plessasb]}. Considering point-like quarks is not sufficient 
to reproduce the observed form factors. In particular, the model eigenfunctions
do not permit contributions from two-body currents arising from pion
exchanges and the semi-relativistic one-body current alone fails to produce
the correct values of the nucleon magnetic moments. However, with the
inclusion of suitable electromagnetic form factors for quarks and considering
an anomalous quark magnetic moment a rather satisfactory agreement with data
is obtained.

Possible improvements of the dynamic model are obviously under discussion.
The pseudoscalar tensor term has been neglected in the model Hamiltonian of
refs.~\cite{[Plessas],[Plessasb]}. 
In ref.~\cite{[Simula]} its effect on the light-baryon
spectrum was estimated together with that of the Thomas-Fermi precession
spin-orbit contribution arising from the scalar confining interaction. The
result was that the agreement with the observed spectra was destroyed. 
The inclusion of vector- ($\rho, \omega,\phi, K^*$) and scalar-meson ($\sigma$)
exchanges was considered in ref.~\cite{[Plessasc]}. 
The tensor forces of vector- and pseudoscalar-meson-exchange interactions 
have opposite signs and largely cancel each other. The effects of the spin-orbit 
forces from the vector- and scalar-meson-exchange interactions are rather weak.
The problem of the spin-orbit force from the Thomas-Fermi
precession (which was not taken into account in ref.~\cite{[Plessasc]}) remains,
however.
In any case, pseudoscalar tensor and
vector-meson exchange contributions modify the model eigenfunctions, so that
one can expect that the two-body currents will contribute even in the simplest
case of the nucleon electromagnetic form factors. Work along these lines is in
progress.

\end{section}

\bigskip

\noindent{\bf Acknowledgements}

\medskip

We are grateful to Willi Plessas for useful discussions and his warm
hospitality at the University of Graz where part of this work was done. This
work was partly performed under the contract ERB FMRX-CT-96-0008 within the
frame of the Training and Mobility of Researchers Programme of the Commission
of the European Union.


\clearpage



\begin{figure}[p]
\hspace*{1cm}\psfig{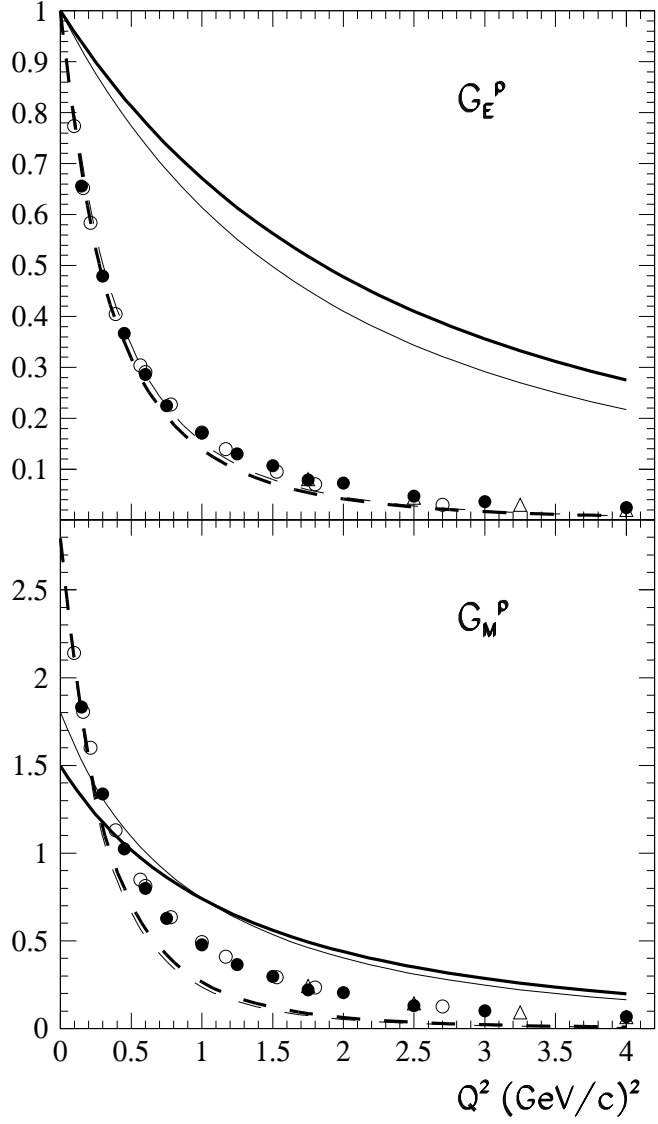}
\vspace{1cm}
\caption{The electric ($G_E^p$) and magnetic ($G_M^p$) form factors  of the
proton as a function of the four-momentum squared $Q^2$. The thin (thick) solid
lines refer to a quark mass $m_{u,d}=$ 250 (340) MeV. 
The thin (thick) dashed lines for $m_{u,d}=250$ (340) MeV include the 
effects of electromagnetic form factors for quarks (see text). 
Experimental points are from ref.~\cite{[Walker]} (solid
circles), ref.~\cite{[Bartel]} (open circles) and ref.~\cite{[Bosted]}
(triangles).}
\end{figure}


\begin{figure}[p]
\hspace*{1cm}\psfig{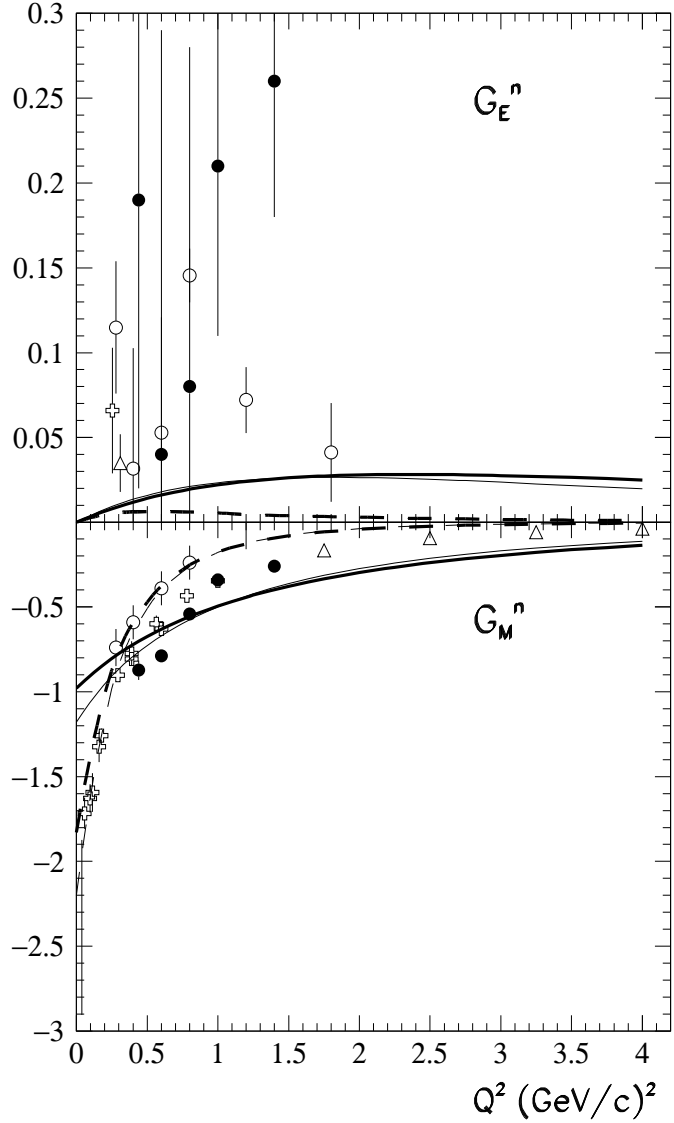}
\vspace{1cm}
\caption{The same as in fig. 1, but for the neutron. Experimental points for
$G_E^n$ are from ref.~\cite{[Hanson]} (open circles), ref.~\cite{[Akerlof]}
(solid circles), ref.~\cite{[Eden]} (cross), ref.~\cite{[Meyerhoff]}
(triangle) and for $G_M^n$ from ref.~\cite{[Bartel]} (crosses),
ref.~\cite{[Bosted]} (triangles), ref.~\cite{[Hanson]} (open circles),
ref.~\cite{[Akerlof]} (solid circles), respectively.}
\end{figure}


\begin{thebibliography}{99}

\bibitem{[IK]}
{N.~Isgur and G.~Karl, Phys. Rev. D18 (1978) 4187; D19 (1979) 2653.}



\bibitem{[Karl]}
{G.~Karl, Int. J. Mod. Phys. E1 (1992) 491.}

\bibitem{[Giannini]}
{M.M.~Giannini, Rep. Progr. Phys. 54 (1990) 453.}

\bibitem{[DeRujula]}
{A.~De R\'ujula, H.~Georgi and S.~L.~Glashow, Phys. Rev. D12 (1975) 147.}

\bibitem{[Godfrey]}
{S.~Godfrey and N.~Isgur, Phys. Rev. D32 (1985) 189.}

\bibitem{[CI]}
{S.~Capstick and N.~Isgur, Phys. Rev. D34 (1986) 2809.}

\bibitem{[Warns]}
{M.~Warns, H.~Schr\"oder, W.~Pfeil and H.~Rollnik, Z. Phys.  C45
(1990) 613, 627.}

\bibitem{[Capstick]}
{S.~Capstick, Phys. Rev.  D46 (1992) 2864.}

\bibitem{[Cardarelli]}
{F.~Cardarelli, E.~Pace, G.~Salm\`e and S.~Simula, Phys. Lett. B357 (1995)
267.}

\bibitem{[Cardarellib]}
{F.~Cardarelli, G.~Salm\`e, S.~Simula and E.~Pace, in {\sl Perspectives in
Hadronic Physics}, eds. S.~Boffi, C.~Ciofi degli Atti and M.M.~Giannini (World
Scientific, Singapore, 1997), p. 403.}

\bibitem{[BT]}
{B.~Bakamjian and L.H.~Thomas, Phys. Rev. 92 (1953) 1300.}

\bibitem{[Cotanch]}
{A.~Szczepaniak, C.-R.~Ji and S.R.~Cotanch, Phys. Rev. C52 (1995) 2738.}

\bibitem{[Manohar]}
{A.~Manohar and H.~Georgi, Nucl. Phys. B234 (1984) 189.}

\bibitem{[Brauer]}
{K.~Br\"auer, A.~Faessler, F.~Fernandez and K.~Shimizu, Nucl. Phys. A507
(1990) 559.}

\bibitem{[Glozman]}
{L. Ya.~Glozman and D.~O.~Riska, Phys. Rep. 268 (1996) 263.}



\bibitem{[Obukhovsky]}
{I.T.~Obukhovsky and A.M.~Kusainov, Phys. Lett. B238 (1990) 142.}


\bibitem{[BHF]}
{A.~Buchmann, E.~Hernandez and A.~Faessler, Phys. Rev C55 (1997) 448.}

\bibitem{[valcarce]}
{A.~Valcarce, P.~Gonz\'ales, F.~Fern\'andez and V.~Vento, Phys. Lett. B367
(1996) 35.}

\bibitem{[fabre]}
{Z.~Dziembowski, M.~Fabre de la Ripelle and G.A.~Miller, Phys. Rev. C53 (1996)
R2038.}

\bibitem{[Shen]}
{P.N.~Shen, Y.B.~Dong, Z.Y.~Zhang, Y.W.~Yu and T.-S.H.~Lee, Phys. Rev. C55
(1997) 2024.}

\bibitem{[Plessas]}
{L.~Ya.~Glozman, Z.~Papp, W.~Plessas, K.~Varga and R.~F.~Wagenbrunn, Phys.
Rev. C57 (1998) 3406.}

\bibitem{[Plessasb]}
{L.~Ya.~Glozman, W.~Plessas, K.~Varga and R.~F.~Wagenbrunn, Phys. Rev. D58 
(1998) 094030.}

\bibitem{[Capuzzi]}
{S.~Boffi, F.~Capuzzi, P.~Demetriou and M.~Radici, preprint
nucl-th/9801066; Nucl. Phys. A637 (1998) 585.}

\bibitem{[Varga]}
{K.~Varga and Y.~Suzuki, Phys. Rev. C52 (1995) 2885; K.~Varga, Y.~Ohbayasi
and Y.~Suzuki, Phys. Lett. B396 (1997) 1.}


\bibitem {[CloseLi]}
{F.~E.~Close and Z.~Li, Phys. Rev.  D42 (1990) 2194.}


\bibitem{[CK]}
{S.~Capstick and B.D.~Keister, Phys. Rev. D51 (1995) 3598.}

\bibitem{[Santopinto]}
{M.~De Sanctis, E.~Santopinto and M.M.~Giannini, Eur. Phys. J. A1 (1998) 187.}

\bibitem{[Vogl]}
{U.~Vogl, M.~Lutz, S.~Klimt and W.~Weise, Nucl. Phys. A516 (1990) 469.}

\bibitem{[Simula]}
{F.~Cardarelli and S.~Simula, preprint hep-ph/9809258.}

\bibitem{[Plessasc]}
{R.~F.~Wagenbrunn, L.~Ya.~Glozman, W.~Plessas, and K.~Varga, 
Few-Body System Suppl. (1998) to be published.}

\bibitem{[Walker]}
{R.C.~Walker {\sl et al.}, Phys. Lett. B 224 (1989) 353.}

\bibitem{[Bartel]}
{W.~Bartel {\sl et al.}, Nucl. Phys. B58 (1973) 429.}

\bibitem{[Bosted]}
{P.~Bosted {\sl et al.}, Phys. Rev. Lett. 68 (1992) 3841.}

\bibitem{[Lung]}
{A.~Lung {\sl et al.}, Phys. Rev. Lett. 70 (1993) 718.}

\bibitem{[Hanson]}
{K.M.~Hanson {\sl et al.}, Phys. Rev. D8 (1973) 753.}

\bibitem{[Akerlof]}
{C.W.~Akerlof {\sl et al.}, Phys. Rev. 135 (1964) B810.}

\bibitem{[Eden]}
{T.~Eden {\sl et al.}, Phys. Rev. C50 (1994) R1749.}

\bibitem{[Meyerhoff]}
{M.~Meyerhoff {\sl et al.}, Phys. Lett. B327 (1994) 201.}

\end{thebibliography}
\end{document}